\documentclass[aps,pra,10pt,twocolumn,showpacs,floatfix,superscriptaddress,noeprint]{revtex4-1}

\usepackage[utf8]{inputenc}
\usepackage[T1]{fontenc}
\usepackage{amsmath,amssymb,amsfonts,bbm,graphicx,times}
\usepackage[dvipsnames]{xcolor}
\usepackage{braket}
\usepackage[normalem]{ulem}
\usepackage[colorlinks=true,bookmarks=false,linkcolor=RoyalBlue,urlcolor=RoyalBlue,citecolor=RoyalBlue,breaklinks]{hyperref}

\newcommand{\effQFI}{ \widetilde{\mathcal{Q}}_{\mathsf{eff}}}
\newcommand{\cssQFI}{\mathcal{Q}_{\mathsf{CSS}}}
\newcommand{\monFI}{\mathcal{F}_{y_t}}
\newcommand{\condQFI}{\bar{\mathcal{Q}}_c}
\newcommand{\rhotilde}{\widetilde{\varrho}}
\DeclareMathOperator{\Tr}{Tr}

\newcommand{\qfi}{\mathcal{Q}}
\begin{document}
\title{Noisy quantum metrology enhanced by continuous nondemolition measurement}
\author{Matteo~A.~C. Rossi}
\affiliation{QTF Centre of Excellence, Turku Centre for Quantum Physics, Department of Physics and Astronomy, University of Turku, FI-20014 Turun Yliopisto, Finland}
\author{Francesco Albarelli}
\affiliation{Faculty of Physics, University of Warsaw, Pasteura 5, 02-093 Warszawa, Poland}
\author{Dario Tamascelli}
\affiliation{Dipartimento di Fisica ``Aldo Pontremoli'', Universit\`a degli Studi di Milano, I-20133 Milano, Italy}
\author{Marco~G. Genoni}
\affiliation{Dipartimento di Fisica ``Aldo Pontremoli'', Universit\`a degli Studi di Milano, I-20133 Milano, Italy}
\affiliation{INFN - Sezione di Milano, I-20133 Milano, Italy}
\begin{abstract}
We show that continuous quantum nondemolition (QND) measurement of an atomic ensemble is able to improve the precision of frequency estimation even in the presence of independent dephasing acting on each atom. We numerically simulate the dynamics of an ensemble with up to $N = 150$ atoms initially prepared in a (classical) spin coherent state, and we show that, thanks to the spin squeezing dynamically generated by the measurement, the information obtainable from the continuous photocurrent scales superclassically with respect to the number of atoms $N$. We provide evidence that such superclassical scaling holds for different values of dephasing and monitoring efficiency. We moreover calculate the extra information obtainable via a final strong measurement on the conditional states generated during the dynamics and show that the corresponding ultimate limit is nearly achieved via a projective measurement of the spin-squeezed collective spin operator.
We also briefly discuss the difference between our protocol and standard estimation schemes, where the state preparation time is neglected.
\end{abstract}
\maketitle
Quantum enhanced metrology~\cite{GiovannettiNatPhot,Braun2018} is one of the most promising and well developed ideas in the realm of quantum technologies, with application ranging from the probing of delicate biological systems~\cite{Taylor2016} to the squeezing enhanced optical interferometry~\cite{Caves1981,RafalPO} recently exploited in gravitational wave detectors~\cite{Acernese2019,Tse2019}.
Atom-based quantum enhanced sensors~\cite{Degen2016,Pezze2018} have also been intensively studied and have myriad of potential applications~\cite{Bongs2019}, most notably in  magnetometry~\cite{Budker2007,Koschorreck2010,Wasilewski2010,Sewell2012,Troiani2018} and atomic clocks~\cite{Ludlow2015,Louchet-Chauvet2010,Kessler2014a}.

Continuous measurements~\cite{WisemanMilburn,Jacobs2014a} have proven to be  very useful tools for the exquisite control of quantum systems, a necessary requirement for the realization of quantum technologies.
The genuinely quantum regime of observing single trajectories has been reached in different platforms, such as superconducting circuits~\cite{Murch2013,Ficheux2018,Minev2019}, optomechanical~\cite{Wieczorek2015,RossiOpto2019} and hybrid~\cite{Thomas2020} systems.
Crucially, continuously monitoring a quantum system allows for the estimation of its characteristic parameters.
A literature has emerged, discussing both practical estimation strategies~\cite{Mabuchi1996,Gambetta2001,Geremia2003,Tsang2010,Cook2014,Six2015,KiilerichHomodyne,Cortez2017,Ralph2017,Atalaya2017,Shankar2019} and the fundamental statistical tools to assess the achievable precision~\cite{Guta2007a,Tsang2011,Tsang2013a,GammelmarkCRB,GammelmarkQCRB,Guta2016,Genoni2017,Albarelli2017a}.

Being also particularly robust against noise~\cite{UlamOrgikh2001}, spin squeezing~\cite{Kitagawa1993,Ma2011} of atomic ensembles has been long studied as a resource for quantum enhanced metrology.
Implementing a continuous quantum nondemolition (QND) measurement of a collective spin observable is a well known approach to generate a conditional spin-squeezed state and the prototypical realization of such schemes relies on the collective interaction between light and atoms~\cite{Takahashi1999,Kuzmich2000,Thomsen2002,Hammerer2010}.
Several measurement-based schemes have been experimentally realized on large atomic ensembles, witnessing spin squeezing of up to $N{ \approx }10^{11}$ atoms~\cite{Appel2009,Schleier-Smith2010,Shah2010,Bohnet2014,Behbood2014,Vasilakis2015,Hosten2016,Bao2020}.

In the ideal noiseless scenario, continuous QND measurements allow one to overcome projection noise and to achieve estimation with Heisenberg limited uncertainty, i.e. inversely proportional to the number of atoms, just by processing the continuous detected signal~\cite{Geremia2003,Molmer2004,Madsen2004,Stockton2004,Chase2009a,Albarelli2017a}.
In conventional metrological schemes exploiting an initial entangled state, Heisenberg scaling is lost in the presence of most kinds of independent noises~\cite{Huelga97,Fujiwara2008,Escher2011,Demkowicz-Dobrzanski2012}.
If the external degrees of freedom causing the noise can be continuously observed, however, its effect can be (at least partially) counteracted and the usefulness of the initial entangled state preserved~\cite{Gefen2016,Plenio2016,Albarelli2018Quantum,AlbarelliIJQI2020}.
The effect of independent noises on continuous QND strategies, in which the entanglement is created dynamically, has not been explored and it will be the main focus of this work.
In more detail, we have the following goals: (i) verify if an enhancement is still observed comparing to the situation where no continuous monitoring is performed; (ii) verify if a quantum enhancement due to non-classical correlations such as spin squeezing and entanglement can still be observed.

\emph{Quantum  metrology  via  continuous QND  monitoring in the presence of dephasing.}
We consider the following scenario: an ensemble of $N$ two-level atoms (qubits) is rotating around the $z$-axis with angular frequency $\omega$; each atom is subjected to equal and independent Markovian dephasing with rate~$\kappa$, leading to the following Lindblad master equation
\begin{align}\label{eq:ME}
\frac{d\varrho}{dt} = \mathcal{L}\varrho \equiv -i \omega [J_z,\varrho] + \frac{\kappa}{2}\sum_{j=1}^N \mathcal{D}[\sigma_z^{(j)}]\varrho \,,
\end{align}
where $J_z = \sum_{j=1}^N \sigma_z^{(j)}/2$, $\mathcal{D}[A]\varrho = A\varrho A^\dag - (A^\dag A \varrho + \varrho A^\dag A)/2$.
Our aim is the estimation of the frequency $\omega$, which in optical magnetometry corresponds to the Larmor frequency $\omega = \gamma B$ ($\gamma$ being the gyromagnetic ratio), thus equivalent to the estimation of the intensity $B$ of a magnetic field.

For noisy quantum frequency estimation schemes, the ultimate limit on the estimation uncertainty $\delta\omega$ for an experiment of total duration $T$, optimized over the duration  $t$ of a single experiment repeated $M=T/t$ times, is given by a quantum Cram\'er-Rao bound (CRB) of the form~\cite{Huelga97}
\begin{align}
(\delta \omega^2) T \geq \frac{1}{\max_t[\mathcal{Q} / t] } \label{eq:QCRB} \,,
\end{align}
 where $\mathcal{Q}$ corresponds to the quantum Fisher information (QFI) of the quantum state evolved up to time $t$ (see Supplemental Material~\cite{sm} for more details on estimation theory~\cite{HelstromBook,Holevo2011b,Hayashi2005,CavesBraunstein,MatteoIJQI}).
\nocite{Dorner2012a,CavesBraunstein,HelstromBook,Holevo2011b,Hayashi2005,CavesBraunstein,MatteoIJQI,Barndorff-Nielsen2000,Rossi2016a,Tamascelli2020,Hayashi2005,Catana2014,Albarelli2017a,GammelmarkQCRB,Macieszczak2016,Guta2016,Mabuchi1996,WisemanMilburn,GammelmarkCRB,Albarelli2018Quantum,Rouchon2015,SteckJacobs,Thomsen2002,Thomsen2002a,Shankar2019,Molmer2004,Hammerer2010,Deutsch2010a,Gough2007,Bouten2007,julia,QContinuousMeasurement,ShammahPIQS,QuTiP2012,noisyqmetrodata,de_Falco_2013,Agarwal1981}

If the initial state is prepared in a coherent spin state (CSS), i.e. the tensor products of eigenstates of the single atom Pauli matrices $\sigma_x^{(j)}$, $|\psi_{\sf CSS} \rangle = \bigotimes_{j=1}^N (|0\rangle_j + |1\rangle_j)/\sqrt{2}$, the state remains separable at all times.
The QFI of the CSS state, optimized over the monitoring time $t$, follows the \emph{standard quantum limit} (SQL), i.e. it is linear in $N$ (corresponding to $\delta\omega \sim 1/\sqrt{N}$ for the uncertainty) and reads
\begin{align}
\mathcal{Q}^\star_{\sf CSS} \equiv {\max_t[\mathcal{Q}_{\sf CSS} / t] } = \frac{N}{2 e \kappa} \,. \label{eq:boundCSS}
\end{align}
By allowing initial entangled states, such as a GHZ state $|\psi_{\sf GHZ}\rangle = ( \otimes_{j=1}^N |0\rangle_j + \otimes_{j=1}^N |1\rangle_j )/\sqrt{2}$, one can achieve a Heisenberg scaling of the QFI, i.e. $\mathcal{Q} \sim N^2$  in the noiseless scenario ($\kappa=0$).

This quantum enhancement is however lost as soon as some non-zero dephasing acts on the system \cite{Huelga97,Fujiwara2008,Escher2011,Demkowicz-Dobrzanski2012}.
Dephasing is the most detrimental among independent noise channels:
remarkably enough, the change of scaling is observed not only asymptotically, but also at finite $N$~\cite{Huelga97}, and most of the approaches suggested in the literature to circumvent the no-go theorems for noisy quantum metrology are useful only in the presence of noise transverse to the Hamiltonian~\cite{Chaves13,Brask15,Sekatski2017metrologyfulland,DemkowiczPRX2017,Zhou2018} or for time-correlated dephasing~\cite{Chin12,Smirne2015a}.

\begin{figure}
\includegraphics[width=0.95\columnwidth]{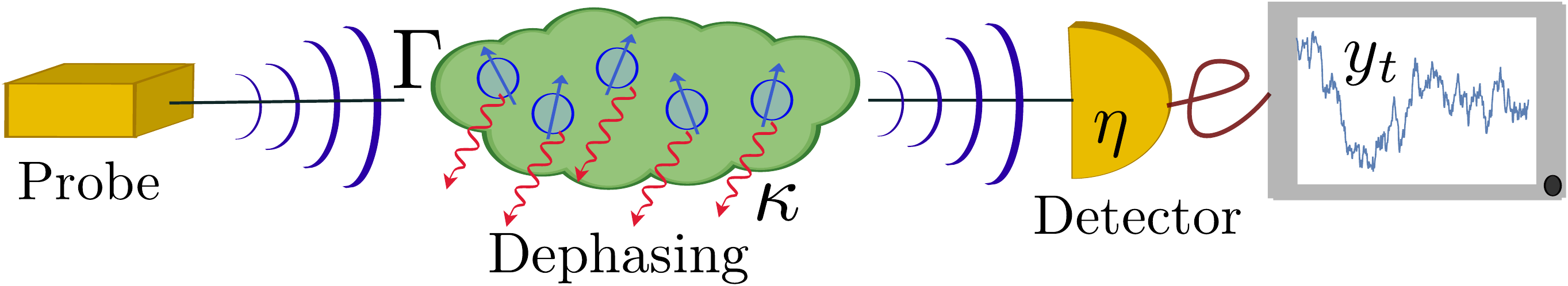}
\caption{\emph{Quantum magnetometry via continuous measurements}: an atomic ensemble of $N$ atoms is sensing a magnetic field that causes precession of the spin around the $z$-axis, and is subjected to independent dephasing on each atom with strength $\kappa$. A far-detuned laser shines the atoms, collectively coupling to the total spin $J_y$ with a strength $\Gamma$, and it is then measured continuously with efficiency $\eta$.
\label{f:setup}}
\end{figure}

We now assume to prepare the atoms in a CSS state $|\psi_{\sf CSS}\rangle$ at time $t=0$, and to perform a continuous monitoring of the collective spin operator $J_y=\sum_{j=1}^N \sigma_y^{(j)}/2$, such that the conditional dynamics of the atom ensemble is described by the stochastic master equation (SME)
\begin{align}
d\varrho_c = \mathcal{L}\varrho_c \,dt + \Gamma\, \mathcal{D}[J_y]\varrho_c \,dt + \sqrt{\eta\Gamma}\, \mathcal{H}[J_y]\varrho_c \,dw \,, \label{eq:SME}
\end{align}
conditioned by the measured photo-current
\begin{align}
\label{eq:photocurr}
dy_t =2 \sqrt{\eta\Gamma} \,\Tr[\varrho_c J_y ] \,dt + dw \,.
\end{align}
The parameter $\Gamma$ corresponds to the collective measurement strength, $\eta$ to the measurement efficiency, $dw$ to a Wiener increment (s.t. $dw^2=dt$) and we have introduced the superoperator $\mathcal{H}[A]\varrho = A\varrho + \varrho A^{\dag} - \Tr[\varrho(A+A^\dag)]\varrho$.
This conditional dynamics can be obtained for instance by considering the setup depicted in Fig.~\ref{f:setup}: a laser is collectively coupled to the total spin of the atoms (possibly inside a cavity) and the outcoming light is continuously measured after the interaction~\cite{Thomsen2002,Madsen2004,Molmer2004,Bouten2007,Nielsen2008a} (more details on these physical implementations are given in the Supplemental Material~\cite{sm}).

%%%%%%%%%%%%
%%%%%%%%%%%%
When one considers these estimation strategies based on continuous measurements, with a dynamics obeying a SME such as Eq.~\eqref{eq:SME}, the parameter can be inferred from two sources of information: the continuous photo-current $dy_t$ and a final strong measurement on the conditional state $\varrho_c$.
In this case the QFI $\mathcal{Q}$ in Eq.~\eqref{eq:QCRB} is replaced by the so-called effective QFI~\cite{Albarelli2017a}
\begin{equation}
\effQFI = \monFI + \sum_{\sf traj} p_{\sf traj} \mathcal{Q}[\varrho_c^{(\mathsf{traj})}] \,,
\end{equation}
that is the classical Fisher information (FI) $\monFI$ that quantifies the information obtainable from the continuous photocurrent $dy_t$, plus the average of the QFI of the conditional states $\varrho_c^{(\mathsf{traj})}$ corresponding to the different trajectories (more details in the Supplemental Material~\cite{sm}).
% On the other hand,
Furthermore, one can also consider the situation where the parameter is inferred from the continuous photocurrent $dy_t$ only; in this scenario the appropriate bound is obtained by replacing $\mathcal{Q}$ with $\monFI$.

In the limit of a large number of atoms $N \gg 1$ and with no noise ($\kappa=0$), it has been already demonstrated that, thanks to this measurement strategy, one can estimate the frequency $\omega$ with a Heisenberg-like scaling, despite the initial state being uncorrelated.
The collective monitoring dynamically generates spin squeezing in the conditional states (and thus entanglement between the atoms), allowing one to observe a $N^2$ scaling both for the effective QFI and the classical FI~\cite{Albarelli2017a}.

Furthermore, in this work we consider a much more practical strategy than the one described in~\cite{Albarelli2018Quantum,AlbarelliIJQI2020}.
There, we have shown that the advantage of an initial entangled state can be recovered by monitoring the $N$ independent environments responsible for the dephasing (typically inaccessible, in practice).
Here, not only we consider a classical (separable) initial state, but we perform continuous monitoring on an ancillary quantum system over which we can assume to have full control; this may correspond, for instance, to an optical field, as depicted in Fig.~\ref{f:setup}.

\emph{Results.}
The SME~\eqref{eq:SME} is invariant under permutation of the different atoms.
This symmetry can be exploited to dramatically reduce the dimension of the density operator $\varrho_c$ as described in~\cite{Chase2008,ShammahPIQS}.
By exploiting some dedicated functions of QuTiP~\cite{QuTiP2012,QuTiP2013} introduced in~\cite{ShammahPIQS}, we have developed a code in Julia (available at~\cite{QContinuousMeasurement}) that has allowed us to simulate quantum trajectories solving the SME~\eqref{eq:SME} and to calculate the figures of merit introduced above up to $N=150$ atoms (see Supplemental Material~\cite{sm} for details on the numerics).

Before moving to the noisy case, we mention that we have been able to verify that for $\kappa=0$ the estimation precision follows a Heisenberg scaling, not only in the limit $N\gg 1$, but also for non-asymptotic values of $N$: our numerics show that both the classical FI $\monFI$ and the average QFI $\condQFI = \sum_{\sf traj} p_{\sf traj} \mathcal{Q}[\varrho_c^{(\mathsf{traj})}]$ (and thus their sum $\effQFI$) are quadratic in $N$ (see Supplemental Material~\cite{sm}).

\begin{figure}[t]
\includegraphics{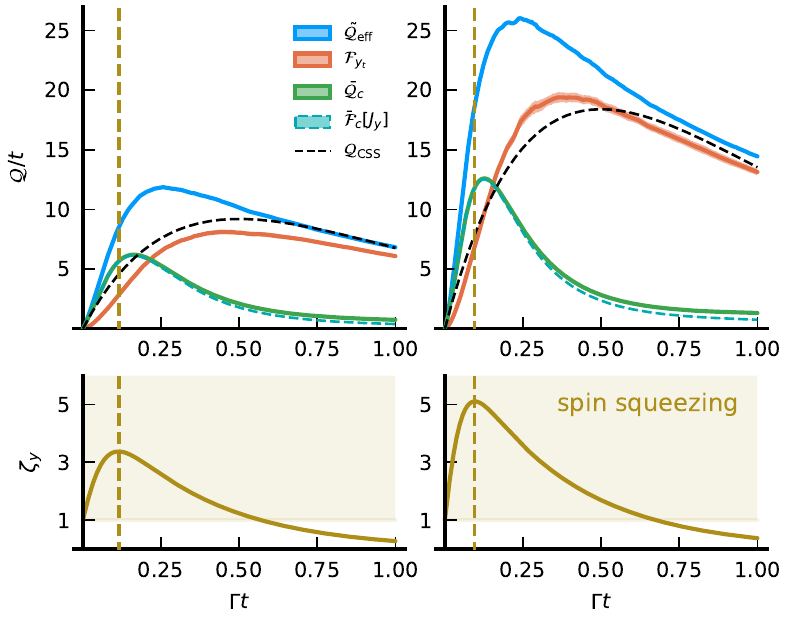}
\caption{Top:
Information rate $\mathcal{Q}/t$ for noisy frequency estimation as a function of time in terms of different figures of merit.
Blue line: effective QFI $\effQFI/t$; orange line: continuous monitoring classical FI $\monFI/t$; green line: conditional states average QFI $\condQFI$; jade green dashed line: conditional states average FI for a $J_y$ measurement $\bar{\mathcal{F}}_c[J_y]$; black dashed line: QFI for a CSS $\cssQFI/t$.
Bottom: average spin squeezing $\bar{\zeta}_y$ as a function of time $\Gamma t$.
Left panels: $N=50$; right panels: $N=100$. The dashed vertical gold line corresponds to the monitoring time where the average spin squeezing violation is maximum.
Parameters: $\kappa/\Gamma=1$, $\omega/\Gamma=10^{-2}$, $\eta=1$, number of trajectories: $n_{\sf traj}=15\,000$. The shaded areas represent the $95\%$ confidence interval (see Supplemental Material~\cite{sm}).}
\label{f:QFIvsTime}
\end{figure}

We now focus on the effect of independent dephasing on this measurement strategy.
In the upper panels of Fig.~\ref{f:QFIvsTime} we plot different figures of merit characterizing our strategy for $N=50$ and $N=100$, comparing them with the results obtained with CSS without monitoring.
We observe that the effective QFI $\effQFI/t$ is larger than the CSS QFI $\cssQFI/t$ at all times.
Remarkably, we observe that for $N=100$ also the maximum of the  monitoring FI $\max_t[\monFI/t]$ surpasses the maximum for the standard strategy $\max_t[\cssQFI/t]$.
In general, this behavior is confirmed for different values of $\kappa$.
This clearly shows that, by increasing $N$, the information obtained from the photocurrent $dy_t$ is enough to achieve a higher precision than via coherent spin states without monitoring.
%%%%% FIGURE %%%%%%%%%%
%%%%%%%%%%%%%%%%%%%%%

We also find that $\condQFI/t$ is larger than $\cssQFI/t$ at certain times.
This result can be explained by studying  the spin-squeezing witness~\cite{Sorensen2001,Thomsen2002,Ma2011} $\zeta_y[\varrho]= \frac{\langle  J_z \rangle^2 + \langle J_x \rangle ^2}{N \Delta J_y^2}$,
where $\langle A \rangle = \Tr[\varrho A]$ and $\Delta J_y^2 = \langle J_y^2\rangle - \langle J_y\rangle^2$.
If $\zeta_y[\varrho] > 1$, the state $\varrho$ is spin-squeezed along the $y$-direction.
In the bottom panels of Fig.~\ref{f:QFIvsTime} we plot the average spin squeezing $\bar{\zeta}_y = \sum_{\sf traj} p_{\sf traj} \zeta_y[\varrho_c^{(\mathsf{traj})}]$ and indeed we observe the maximum violation approximately at the same time $t$ for which $\condQFI/t > \cssQFI/t$ (for more details about the distribution of trajectory dependent quantities, as $\zeta_y[\varrho_c^{(\mathsf{traj})}]$ and $\mathcal{Q}[\varrho_c^{(\mathsf{traj})}]$, see Supplemental Material~\cite{sm}).
The generation of spin-squeezed conditional states leads us to investigate the effectiveness of a strong measurement of the operator $J_y$, optimal in the noiseless case~\cite{Albarelli2017a}.
We evaluate the corresponding classical Fisher information, and we average it over the different trajectories, yielding $\bar{\mathcal{F}}_c[J_y]$.
As shown in Fig.~\ref{f:QFIvsTime}, $\bar{\mathcal{F}}_c[J_y]$ is approximately equal to $\condQFI$ for evolution times near to the maximum of both $\condQFI$ and $\effQFI$.
In general, our numerics show that a strong measurement of $J_y$ is nearly optimal in the parameter regime relevant for our protocol.
%
%%%%%%%%%%
\begin{figure}[t!]
\includegraphics{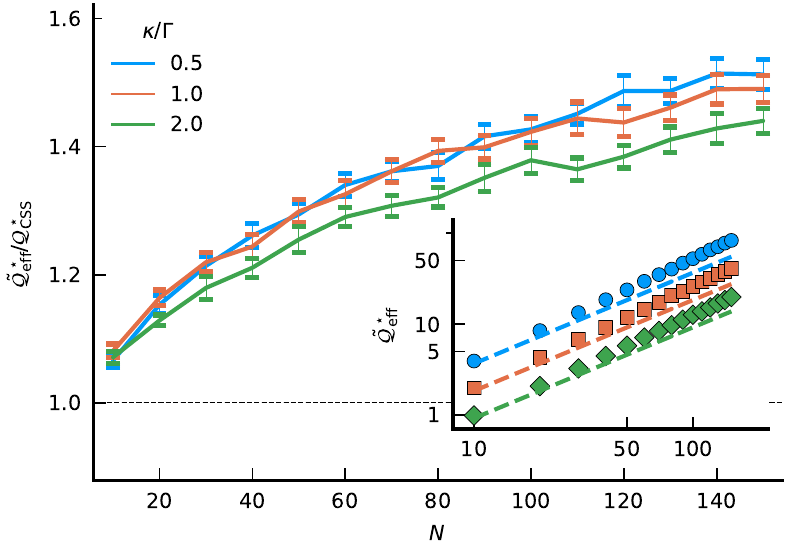}
\caption{Ratio between the optimized effective QFI $\effQFI^\star$ and the optimized $\mathcal{Q}_{\sf CSS}^\star$ (dashed lines) as a function of $N$ for different values of the dephasing rate $\kappa$, with $\omega/\Gamma=10^{-2}$ and $n_{\sf traj}=10\,000$ trajectories.
See the Supplemental Material~\cite{sm} for details on the statistical error. In the inset, log-log plot of $\effQFI^\star$ (markers) and $\mathcal{Q}_{\sf CSS}$  (dashed lines) as a function of $N$ for the same values of $\kappa$.
\label{f:eQFIvsN}}
\end{figure}
%%%%%%

Importantly, the behaviour of the spin-squeezing witness $\bar{\zeta}_y$ helps us also to better understand the optimal monitoring times for the different figures of merit plotted in Fig. \ref{f:QFIvsTime}.
The following relationship holds:
\begin{align}
t_{\sf opt}[\condQFI] < t_{\sf opt}[\effQFI] < t_{\sf opt}[\monFI] \,. \label{eq:optimaltime}
\end{align}
In order to maximize the average QFI $\condQFI/t$, as we discussed above, one therefore needs to stop the monitoring at a time $t_{\sf opt}[\condQFI]$ corresponding approximately to the maximum spin squeezing.
On the other hand, since $\monFI$ quantifies the information contained in the photo-current $y_t$ accumulated during the whole monitoring time, one can fully exploit the generated spin squeezing and the encoding of the parameter by waiting longer, i.e. $t_{\sf opt}[\monFI] > t_{\sf opt}[\condQFI]$.
Consequently, since the effective QFI $\effQFI$ is the sum of $\monFI$ and $\condQFI$, the corresponding optimal time has to satisfy the relation in~\eqref{eq:optimaltime}.

\begin{figure}[t!]
\centering
\includegraphics{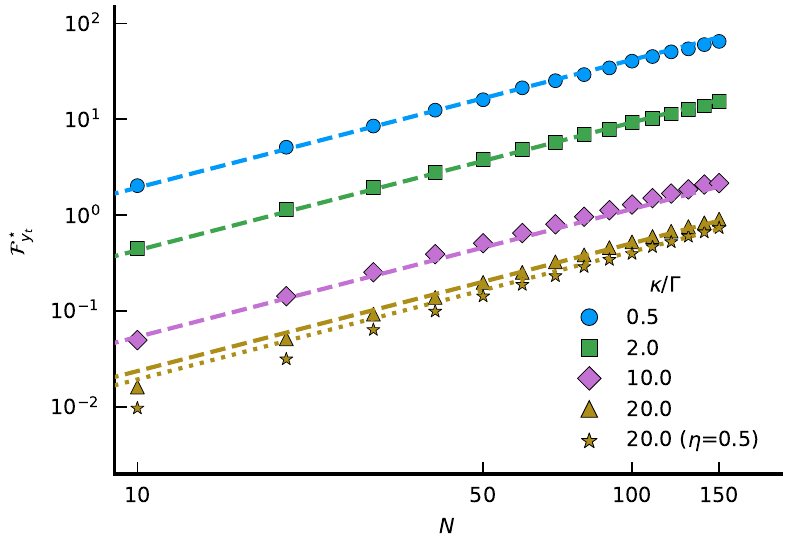}
\caption{Continuous monitoring FI $\max_t[\monFI/t]$ (markers) as a function of $N$ for different values of the dephasing rate, with $\omega/\Gamma=10^{-2}$ and number of trajectories $n_{\sf traj}=10\,000$. Dashed lines showing super-linear functions scaling as $N^{4/3}$ have been plotted as a guide to the eye.
\label{f:monFIvsN}}
\end{figure}

Fig.~\ref{f:eQFIvsN} shows the ratio between the optimized effective QFI $\effQFI^\star \equiv \max_t [ \effQFI / t]$ and the CSS bound $\cssQFI^*$ as a function of $N$ and for different values of the dephasing rate $\kappa$.
It is clear from the plot and from the inset, where the two quantities are plotted in logarithmic scale, that not only the CSS bound is always surpassed, but also the effective QFI shows a super-linear behavior.
An important role in this result is played by the photo-current FI $\monFI$, which corresponds to the most practical strategy of estimating $\omega$ without any strong final measurement.
As it is apparent from Fig.~\ref{f:monFIvsN}, the behavior of $\monFI^\star \equiv \max_t[\monFI/t]$ is very peculiar: a $\kappa$-independent super-classical scaling $\monFI^\star \sim N^{4/3}$
seems to hold for all the considered values of the dephasing strength $\kappa$ (notice that by increasing $\kappa$ the scaling $N^{4/3}$ is obtained and then maintained for large enough $N$).
It is also important to mention that a reduced measurement efficiency (e.g. $\eta=0.5$ in one of the curves in Fig.~\ref{f:monFIvsN}) yields the same qualitative results as having a larger dephasing:  (i) the CSS QFI is surpassed as long as $N$ is large enough; (ii) despite non-unit efficiency, the $\kappa$-independent scaling $N^{4/3}$ is still observed for $\monFI$, but for larger $N$ and with a reduced proportionality constant (more plots for $\eta<1$ are found in the Supplemental Material~\cite{sm}).

Finally, we consider the performance of our strategy in the presence of collective Markovian dephasing, that is, a dynamics described by a master equation as in Eq.~\eqref{eq:ME}, but with the last term replaced by $\kappa_{\sf coll}  \mathcal{D}[J_z]\varrho$.
Also in this case our scheme based on continuous QND monitoring performs better than a standard strategy with CSS states and no monitoring.
However, we observe that spin-squeezing is hardly generated and that no enhancement in the estimation precision due to quantum correlations can be observed (see more details in the Supplemental Material~\cite{sm}).
It is crucial to remark that collective dephasing is best tackled with specific estimation protocols, exploiting decoherence-free subspaces, that are able to restore Heisenberg scaling~\cite{Dorner2012a}.
We therefore leave to future investigations the possibility of combining these strategies with our approach, to jointly counteract both independent and collective dephasing.

%%%%%%%%%%%

\emph{Discussion.}
We showed that continuous QND monitoring leads to an enhancement in the estimation precision, even in the presence of Markovian dephasing, known to be the most detrimental noise for quantum metrology.

One last remark regarding the precision our protocol can ultimately achieve is in order.
A fundamental bound that covers strategies with ancillary systems and full and fast control~\cite{Sekatski2017metrologyfulland,DemkowiczPRX2017} shows that only an improvement of a factor $e$ on $\mathcal{Q}^\star_{\sf CSS}$ in Eq.~\eqref{eq:boundCSS} can be obtained, i.e. $\mathcal{B}_{\sf ent} = N/(2\kappa)$.
This bound is attained asymptotically for $N \gg 1 $ by preparing a spin squeezed initial state, without ancillas and control operations~\cite{UlamOrgikh2001}.
At present it is not clear if the effective QFI for our scheme should also obey this bound.
Continuous monitoring can be described as qubits interacting with the system and being sequentially measured~\cite{Brun2002,CiccarelloQMETR,Gross2017}.
However, it is unclear if the assumptions beyond the derivation in~\cite{Sekatski2017metrologyfulland,DemkowiczPRX2017} are satisfied in the limit of infinitesimal time steps with simultaneous enconding, noise and interaction with the ancillas.
Despite the high optimization level of our code, we could not investigate regimes where our strategy would be able to reach values near to $\mathcal{B}_{\sf ent}$.
We thus leave as an open question if our protocol, thanks in particular to the observed scaling of the classical FI $\monFI^*$, may be able to attain (or possibly surpass) this bound  in experimentally relevant regimes (state of the art experiments with atomic clouds involve $10^{5}$ -- $10^{11}$ atoms~\cite{Vasilakis2015,Hosten2016,Bao2020}).

Finally, we highlight again one of the main features of our protocol: the monitoring-induced dynamics generates the resourceful state simultaneously with the frequency encoding.
In fact, in the standard analysis of quantum estimation strategies the state preparation time is typically neglected.
A fair comparison between ``classical'' and ``quantum enhanced'' strategies accounting also the preparation time as a resource is discussed, for the noiseless scenario, in~\cite{Dooley2016,Hayes2018,Haine2020}.
In~\cite{Hayes2018,Haine2020}, in particular, the generation of spin squeezing via one-axis and two-axis twisting is considered and it is shown that the best strategy is to allow the encoding and the spin-squeezing Hamiltonians to act simultaneously.
Remarkably enough, this enhancement is comparable to the one we observe in our protocol, with no need of time-dependent control Hamiltonians.

The role of preparation time in noisy metrology with Markovian independent dephasing has been discussed in~\cite{Dooley2016}.
There, however, only initial GHZ states have been considered and, unsurprisingly, they offer no improvement over CSS states;
the same is true when the preparation time is not taken into account~\cite{Huelga97}.
Optimal entangled states for standard frequency estimation in the presence of dephasing have been numerically obtained in~\cite{Frowis2014}.
We observe that, remarkably, our protocol can achieve an enhancement of the same order of magnitude (cf. Fig.~\ref{f:eQFIvsN} and Fig. 3(b) of~\cite{Frowis2014}).
We therefore expect that, if the preparation time is counted as a resource, our protocol should be able to outperform the one involving the preparation of those optimal states.

Concluding, our results pave the way to further theoretical and experimental investigations into noisy quantum metrology via QND continuous monitoring,
as a practical and relevant tool to obtain a quantum enhancement in spite of decoherence.

\begin{acknowledgments}
{\em Acknowledgments}.
We thank N.~Shammah for several helpful discussions and suggestions regarding QuTiP and its library for permutationally invariant systems~\cite{ShammahPIQS}.
We also thank J.~Amoros~Binefa, R.~Demkowicz-Dobrzański, J.~Kołodyński and A.~Smirne for helpful discussions.
MACR acknowledges financial support from the Academy of Finland via the Centre of Excellence program (Project no. 312058).
FA acknowledges financial support from the National Science Center (Poland) grant No. 2016/22/E/ST2/00559.
MGG acknowledges support from a Rita Levi-Montalcini fellowship of MIUR.
MGG and DT acknowledge support from the Sviluppo UniMi 2018 initiative.
The computer resources of the Finnish IT Center for Science (CSC) and the FGCI project (Finland) are acknowledged.
\end{acknowledgments}

\bibliography{library}

\clearpage

\appendix
\setcounter{figure}{0}
\setcounter{page}{1}
\renewcommand{\thefigure}{S\arabic{figure}}
\renewcommand{\theequation}{S\arabic{equation}}

%% For prl style (but no numbers in appendices)
% \setcounter{equation}{0}
% \setcounter{figure}{0}
% \setcounter{table}{0}
% \setcounter{page}{1}
% \makeatletter
% \renewcommand{\theequation}{S\arabic{equation}}
% \renewcommand{\thefigure}{S\arabic{figure}}

\begin{center}
	\textbf{\large Supplemental material: Noisy quantum metrology enhanced by continuous nondemolition measurement}
\end{center}

The structure of this Supplemental Material is as follows: in Sec.~\ref{a:qet} we present a more detailed discussion of quantum estimation theory for continuous measurements, introducing the quantities defined in the main text; in Sec.~\ref{a:phys} we elaborate on the physical meaning of the stochastic master equation (SME) we consider in this work; in Sec.~\ref{a:code} we discuss the implementation of the numerical solution of the dynamics and on its computational complexity, and we discuss the statistical error analysis; in Sec.~\ref{a:noiseless_finiteN} we show the results in absence of dephasing noise; in Sec.~\ref{a:conditional_states} we show some additional properties of the conditional states generated during the dynamics, by studying the distributions of the corresponding spin squeezing and quantum Fisher information, and by presenting a specific realization of the conditional state and visualizing its evolution on the Bloch sphere; in Sec.~\ref{a:finite_efficiency} we present results for finite measurement efficiency and finally in Sec.~\ref{a:collective_dephasing} we analyze the case of collective dephasing noise.

\section{Quantum estimation theory for continuous measurements}
\label{a:qet}

We start from the classical problem of estimating the true value of a parameter $\omega$ that enters into the conditional probability $p(x|\omega)$ of observing the measurement outcome $x$.
Under very general assumptions, the uncertainty (quantified by the root mean square error) of any unbiased estimator is lower bounded by the classical Cramér-Rao bound (CRB), as follows
\begin{equation}
\delta \omega \geq \frac{1}{\sqrt{M \mathcal{F}[p(x|\omega)]}} \,,
\end{equation}
where $M$ is the number of measurements performed and $\mathcal{F}[p(x|\omega)] =
\sum_x p( x|\omega ) \left[\partial_\omega \ln p( x|\omega ) \right]^2$ is the classical Fisher information (FI).

When dealing with quantum systems, probabilities densities are obtained from the Born rule $p(x|\omega)=\Tr[\varrho_\omega \Pi_x]$, where $\varrho_\omega$ is a family of quantum states parametrized by $\omega$, and $\{ \Pi_x \}$ is a positive-operator valued measure (POVM) describing the statistical properties of the measurement apparatus.
By optimizing over all POVMs one obtains the quantum CRB~\cite{HelstromBook,Holevo2011b,Hayashi2005,MatteoIJQI,CavesBraunstein}
\begin{align}
\delta \omega \geq \frac{1}{\sqrt{M \mathcal{F}[p(x|\omega)]}} \geq \frac{1}{\sqrt{M \qfi[\varrho_\omega]}}\,, \label{eq:QCRB_2}
\end{align}
where
\begin{align}
\qfi[\varrho_\omega] =
%4 \frac{D_B \left[ \varrho_\omega, \varrho_{\omega + d \omega} \right] }{d \omega^2} = \frac{ 8 \left(1 - F \left[ \varrho_\omega, \varrho_{\omega + d \omega}  \right] \right) }{ d \omega^2 }
\lim_{\epsilon \rightarrow 0} \frac{ 8 \left(1 - F \left[ \varrho_\omega, \varrho_{\omega + \epsilon}  \right] \right) }{ \epsilon^2 }
\end{align}
is the quantum Fisher information (QFI), expressed in terms of the fidelity between quantum states $F \left[\varrho_1,\varrho_2 \right] = \Tr \left[ \sqrt{\sqrt{\varrho_1} \varrho_2 \sqrt{\varrho_1} } \right]$ \cite{CavesBraunstein,Rossi2016a,Tamascelli2020}.
The QFI $\qfi[\varrho_\omega]$ depends only on the local properties of the family of states around the true value of $\omega$, and the optimal POVM that satisfies $\mathcal{F}[p(x|\omega)]=\qfi[\varrho_\omega]$ always exists (but a two-step strategy~\cite{Barndorff-Nielsen2000,Hayashi2005} might be needed to actually saturate the bound).

In this work we consider a quantum state evolving according to the stochastic master equation (SME) introduced in Eq.~(4) of the main text,
\begin{align}
d\varrho_c = \mathcal{L}\varrho_c \,dt + \Gamma\, \mathcal{D}[J_y]\varrho_c \,dt + \sqrt{\eta\Gamma}\, \mathcal{H}[J_y]\varrho_c \,dw \,, \label{eq:SME}
\end{align}
conditioned on the observed photocurrent $y_t$, which is a stochastic process defined by
\begin{align}
\label{eq:photocurr}
dy_t =2 \sqrt{\eta\Gamma} \,\Tr[\varrho_c J_y ] \,dt + dw \,
\end{align}
depending on $\omega$ through the expectation value of $J_y$.

As explained in the main text, we can extract information on $\omega$ both from the current $y_t$ and by measuring the conditional state $\varrho_c$.
Once the SME (and thus the monitoring choice) is fixed, the correct CRB is written in terms of an effective QFI~\cite{Catana2014,Albarelli2017a}
\begin{align}
\widetilde{\qfi}  = \mathcal{F}[p_\mathsf{traj}] + \sum_\mathsf{traj} p_\mathsf{traj} \qfi[\varrho_c^{(\mathsf{traj})}] \,, \label{eq:effQFI}
\end{align}
where informally the sum over trajectories represents the integration over all the realizations of the stochastic process $y_t$, $p_\mathsf{traj}$ represents the probability density corresponding to a particular realization and $\varrho_c$ the solution of the SME corresponding to the same realization.
In other words, $\widetilde{\qfi}$ is equal to the sum of the classical FI $\mathcal{F}[p_\mathsf{traj}]=\sum_{\mathsf{traj}} \frac{ (\partial_\omega p_\mathsf{traj} )^2 }{p_\mathsf{traj}}$, plus the average of the QFIs of the corresponding conditional states $\qfi[\varrho_c]$.
Physically, the first term quantifies the amount of information obtained by continuous measurement of the light that has interacted collectively with the atoms.
The second term quantifies the maximal amount of information that can be obtained by stopping the conditional evolution and performing a (strong) measurement on the resulting conditional state.

We mention that a more general bound, which only depends on the interaction between the radiation and the atoms but is optimized over the measurement strategy on the outgoing radiation was presented~\cite{GammelmarkQCRB,Macieszczak2016,Guta2016}.
Unfortunately, this type of ultimate bound is not meaningful when the dynamics includes unmonitored noise channels, such as the independent dephasing considered in this work.

The quantity $\mathcal{F}[p_\mathsf{traj}]$ can in principle be obtained by considering a linear version of the SME
\begin{align}
d \tilde{\varrho}_c = \mathcal{L}\tilde{\varrho}_c \,dt + \Gamma\, \mathcal{D}[J_y]\tilde{\varrho}_c \,dt + \sqrt{\eta\Gamma}\, ( J_y\tilde{\varrho}_c + \tilde{\varrho}_c J_y ) \,dw \,, \label{eq:SMElin}
\end{align}
since the quantity $\Tr \tilde{\varrho}_c$ corresponds to $p_\mathsf{traj}$~\cite{Mabuchi1996,WisemanMilburn} (up to a parameter-independent proportionality constant).

A practical method to evaluate this FI numerically was proposed in~\cite{GammelmarkCRB} and instead of solving~\eqref{eq:SMElin} it requires solving the original SME~\eqref{eq:SME} plus an additional stochastic equation coupled to it.
In our previous paper~\cite{Albarelli2018Quantum} we have presented a concrete implementation of this method that takes advantage of a stable and effective method to solve SMEs numerically~\cite{Rouchon2015}, and that also allows to evaluate efficiently the QFI of the conditional states $\mathcal{Q}[\varrho_c]$ and thus the full effective QFI $\effQFI$. In the following section we give more details about the numerical implementation.

\section{Physical implementations}\label{a:phys}

In this section we give a brief overview of two different physical setups that are described by the same SME (4), providing a more precise explanation of the schematic representation of Fig.~1.

The abstract idea is to engineer a quantum non-demolition (QND) interaction of the form $H_\text{QND} \propto J_y p $, where $J_y$ is a collective spin operator of the atoms and $p$ is the ``momentum'' operator of an continuous variable ``meter'' system, initially prepared in a pure Gaussian state.
In the limit of a very strong QND interaction, measuring the conjugate ``position'' observable $x$ after the interaction effectively implements an instantaneous (``strong'') measurement of $J_y$.
On the other hand, in the context of continuous measurements the interaction is assumed to be very weak, but the ``meter'' is a continuous train of modes on which $x$ is continually measured, see e.g.~\cite{SteckJacobs}.
Continuous measurements are not instantaneous and thus there is a nontrivial interplay between the free dynamics of the system and the the measurement-induced backaction that tends to collapse the state to an eigenstate of $J_y$.

The proposals to implement QND measurements on atomic ensembles rely on coherent light-matter interactions, even if the physical meaning of the operator $p$ appearing in $H_\text{QND}$ can be different.
Anyhow, one has to assume a uniform coupling of the light with all the $2J=N$ two-level atoms so that the dynamics can be described by collective spin $J$ operators as in $H_\text{QND}$ and in Eq.~(4).
The motional degrees of freedom of the atoms are neglected in this analysis.

In the first setting, the atoms are placed inside a leaky optical cavity, with a resonant frequency far detuned from the splitting of the two relevant atomic levels.
This can be described as a dispersive interaction that induces a phase shift of the cavity field, proportional to the atom number difference between the two levels.
The cavity is driven by a strong classical field and the light leaking out of the cavity is continuously measured with a homodyne detector.

Assuming a bad-cavity regime, i.e. in the presence of a large cavity decay rate, the cavity mode can be adiabatically eliminated, so that the sought QND interaction between the atoms and the traveling light is effectively implemented and the SME~(4) describes the dynamics~\cite{Thomsen2002,Thomsen2002a}.
% The ``measurement strenght'' parameter $\Gamma$ will depend on the detuning $\Delta$ of the cavity field, on the

While such cavity-based proposals were initially focused on inducing spin-squeezing via QND measurements, it is indeed also possible to include a field to be sensed, perpendicular to the driving, to fully reproduce our Eq.~(4).
A different cavity based setup requiring two driving lasers has been recently proposed and experimentally implemented in~\cite{Shankar2019}.
% \FA{Non sono sicurissimo di aver capito bene questa cosa, nel senso che non capisco se la SME di Shankar è quella di Thomsen o se stanno usando un effetto fisico diverso... è che in Shankar hanno 2 driving...}

A similar analysis can be applied to atoms in free space, ending up with the same dynamics, but with a different expression for the  ``measurement strength'' parameter $\Gamma$~\cite{Thomsen2002a}.
However, for atomic ensembles probed by traveling light modes, it is more common to take advantage of the magneto-optical Faraday effect to implement a QND measurement~\cite{Molmer2004,Hammerer2010,Deutsch2010a}.
The Faraday interaction of the atoms with a far-off-resonant laser is effectively a QND interaction between the collective spin of the ensemble and the Stokes operator of the light, which takes the role of $p$.

In this setup a linearly polarized optical probe is transmitted through the atomic cloud.
Due to the Faraday interaction the two circular components of the polarization have a different coupling with the excited state and thus the probe field acquires a phase proportional to the population difference between the two levels.

The continuous measurement is performed by a balanced polarimeter, consisting of a polarizing beam splitter and a differential photo detector at the two outputs.
This measurement scheme is physically different from homodyne detection, but has the same formal description (treating the Stokes operators perpendicular to the direction of propagation of the beam as the optical quadratures) and the same SME~(4) can thus be obtained.

Finally, we remark that a rigorous mathematical treatment of the adiabatic elimination, both of the cavity~\cite{Gough2007} and of the excited states~\cite{Bouten2007} is not trivial, but the usual expected results are obtained in the appropriate regimes.

\section{Details on the numerical implementation}
\label{a:code}
The code used to obtain the results presented in this manuscript is written in Julia \cite{julia} and is available on Github.com \cite{QContinuousMeasurement}.

In a nutshell, the code implements the algorithm described in \cite{Albarelli2018Quantum} which consists in the Montecarlo generation of the two solutions of two coupled SMEs, which can be written as:
\begin{widetext}
\begin{align}
    \rho_{t+dt} & = \rhotilde_{t+dt} / \Tr[\rhotilde_{t+dt}], \qquad \text{where}\qquad  \rhotilde_{t+dt} = M_{dy} \rho_t M_{dy}^\dag + (1-\eta) \Gamma J_y \rho_t J_y^\dag dt + \frac{\kappa}{2} \sum_{j=1}^N \sigma_z^{(j)} \rho_t \sigma_z^{(j)\dag} dt \label{eq:rho_numerical} \\
    \tau_{t+dt} & = \Tr[\rhotilde_{t+dt}]^{-1} \left[ \partial_\omega M_{dy} \rho_t M_{dy}^\dag + M_{dy} \tau_t M_{dy}^\dag + M_{dy} \rho_t (\partial_\omega M^\dag_{dy} ) + (1-\eta)\Gamma J_y \tau_t J_y^\dag dt + \frac{\kappa}{2} \sum_{j=1}^N \sigma_z^{(j)} \tau_t \sigma_z^{(j)\dag} dt \right], \label{eq:tau_numerical}
\end{align}
where we have defined the Kraus operator
\begin{equation}
    M_{dy} = \mathbb{I} - i H_\omega dt - \frac \kappa 4 N \mathbb{I} \, dt - \frac{\Gamma}{2} J_y^2 dt + \sqrt{\eta \Gamma} J_y dy_t + \frac{\eta \Gamma}{2} J_y^2 (dy_t^2 - dt). \label{eq:kraus}
\end{equation}
\end{widetext}
In the third term of Eq.~\eqref{eq:kraus}, we have used $\sum_j \sigma_z^{(j) \dag} \sigma_z^{(j)} = N\mathbb{I}$, while the last term represents the Euler-Milstein correction ($dt$ is not infinitesimal in the numerical integration of the stochastic differential equation and therefore $dy_t^2 \neq dt$~\cite{Rouchon2015}).
The continuous photocurrent $dy_t$ was introduced in Eq.~(5) of the main text:
\begin{equation}
    dy_t = 2 \sqrt{\eta\Gamma} \Tr[J_y\rho_{t+dt}] dt + dw
\end{equation}

From $\rho_{t+dt}$ and $\tau_{t+dt}$ we can evaluate the relevant figures of merit, as discussed in \cite{Albarelli2018Quantum}: $\monFI = \mathbb{E}[(\Tr \tau_t)^2]$ and $\bar{Q}_c(t) = \mathbb{E}[Q[\rho_t]]$, where $\mathbb{E}[\cdot]$ denotes the average over the sampled trajectories and
\begin{equation}\label{eq:qfi}
    \mathcal{Q}[\rho_t] = 2\sum_{\lambda_i + \lambda_j \neq 0} \frac{|\braket{\psi_i|\partial_\omega\rho_t|\psi_j}|^2}{\lambda_i + \lambda_j}
\end{equation}
is the QFI for the state $\mathcal{Q}[\rho_t]$~\cite{MatteoIJQI}.
In Eq. \eqref{eq:qfi}, $\rho_t = \sum_i \lambda_i \ket{\psi_i}\!\bra{\psi_i}$ is the diagonalization of the density matrix and one can obtain $\partial_\omega \rho_t$ from $\rho_t$ and $\tau_t$ with the formula $\partial_\omega \rho_t = \tau_t - \rho_t \Tr \tau_t$ \cite{Albarelli2018Quantum}.

As explained in the main text, we exploit the permutational symmetry of the state by expressing it in the Dicke basis \cite{ShammahPIQS}. Thus, instead of considering the whole Hilbert space, with size growing as $2^N$, we can restrict to a subspace of size proportional to $N^4$. Moreover, the density operator in the Dicke basis has a block-diagonal structure, with blocks of sizes $2j + 1$, where $j$ is the spin number, a (half-)integer number ranging from $N/2$ to $0, 1/2$ for $N$ even or odd. The total number of non-zero elements of $\rho$ is thus of order $N^3$ \cite{ShammahPIQS}, with a consequent advantage in memory consumption. The initial coherent spin state occupies only the first block with $j = N/2$. The other blocks are populated during the dynamics due to the dephasing noise.

All the global spin operators present in Eqs. (\ref{eq:rho_numerical}-\ref{eq:kraus}) can be easily expressed in the Dicke basis, and their action is confined to each subspace at fixed $j$. The last terms of Eqs. \eqref{eq:rho_numerical} and \eqref{eq:tau_numerical}, as they are written, would be computationally heavy to calculate. However, they can be conveniently expressed in the Dicke basis in terms of the Liouvillian $L$ of the superoperator $\sum_j\mathcal{D}[\sigma_z^{(j)}]$, acting on the vectorized density matrix $\vec\rho_t$ (where the columns of the matrix $\rho_t$ are stacked). Specifically,
\begin{equation}
    \frac{\kappa}{2} \sum_j \sigma_z^{(j)} \rho_t \sigma_z^{(j)\dag} dt \longrightarrow \frac \kappa 2 (L+N \mathbb{I})\vec\rho_t,
\end{equation}
and similarly for the last element of Eq. \eqref{eq:tau_numerical}. We obtain the matrix $L$ by employing the PIQS module \cite{ShammahPIQS} of the QuTiP library \cite{QuTiP2012}. All the operators and the matrix $L$ contain a large number of zeros, and hence they are encoded as sparse matrices, for memory efficiency and computational speed, while for $\rho_t$ and $\tau_t$ each block is stored in dense format. The block-diagonal structure of $\rho_t$ and $\tau_t$ is also exploited in the evaluation of the QFI, as the latter is simply the sum of the QFIs for each block (thus requiring the diagonalization of small dense matrices, instead of the full sparse matrix).

As explained in the text, in the case of zero dephasing ($\kappa = 0$), the dynamics is confined to the block $j= N/2$, which has dimension $N+1$, and hence the computational complexity of the simulation is further reduced, allowing us to reach values of $N = 300$ and even further.

Equations (\ref{eq:rho_numerical}-\ref{eq:tau_numerical}) are simulated for a large number $n_{\sf traj}$ of trajectories. By averaging over the trajectories, we build our estimators for $\monFI$ and $\effQFI$. The estimate errors are assessed via standard bootstrapping, by considering $95 \%$ confidence intervals for the deviations from our sample mean. The errors are below $2\%$ for $\monFI$ and below $1.5\%$ for $\condQFI$ for all $t$ for $n_{\sf traj} = 10\,000$. Bootstrapping is also employed to estimate $95 \%$ confidence intervals for the optimal values $\monFI^\star$ and $\effQFI^\star$.
For $n_{\sf traj} = 10\,000$, uncertainties are typically below $2\%$, and are not shown in all the logarithmic plots, as the error bars would be indistinguishable from the markers.

Finally, in order for the simulation to be accurate, the time step $dt$ need to be chosen so that it is much smaller than the characteristic time of the dynamics. We have extensively tested the convergence of the numerical dynamics for decreasing values of $dt$, and we have verified that, for the values of $\omega, \Gamma, \kappa$ and $N$ we have considered, $dt\approx 10^{-4} \div 10^{-5}$ is sufficiently small, with larger values of $N$ requiring smaller values of $dt$.

Thanks to Julia's distributed computing capabilities, the code (available at this link \cite{QContinuousMeasurement}) can be readily run on HPC clusters, with massively parallel simulation of the trajectories. The data used to produce the figures in the manuscript is available at \cite{noisyqmetrodata}.
%%%%%
%%%%%

%%%%%%%%%%%%%
\begin{figure}
    \includegraphics{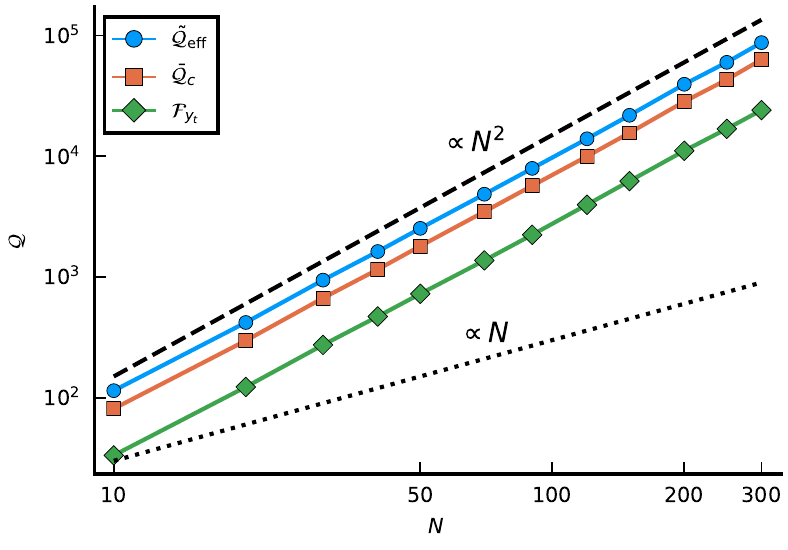}
    \caption{Effective QFI $\effQFI$, monitoring FI $\monFI$ and conditional states average QFI $\condQFI$ for noiseless frequency estimation ($\kappa=0$) as a function of $N$ for $\Gamma t=2$, $\omega/\Gamma=10^{-2}$ and $\eta=1$.
    Linear $\sim N$ (dotted line) and quadratic $\sim N^2$ (dashed line) functions are shown as a guide to the eye.
    The results are obtained from 10\,000 trajectories, the statistical uncertainty is too small to be appreciated (see Sec. \ref{a:code} for details).}
    \label{f:nodephasing}
\end{figure}
%%%%%%%%%%%%

\section{Noiseless results for finite $N$}
\label{a:noiseless_finiteN}

As we mention in the main text, it was already demonstrated that for $\kappa=0$ the estimation precision follows a Heisenberg scaling in the limit $N\gg 1$.
Our numerics show that this scaling is observed also for non-asymptotic values of $N$ as shown in Fig.~\ref{f:nodephasing}.
Both the classical FI $\monFI$ and the average QFI $\condQFI = \sum_{\sf traj} p_{\sf traj} \mathcal{Q}[\varrho_c^{(\mathsf{traj})}]$ (and thus their sum $\effQFI$) are quadratic in $N$.
One should notice that for $\kappa=0$ all the operators entering in the SME~\eqref{eq:SME} are collective operators and therefore one can further reduce the dimension of the relevant Hilbert space to $d=N+1$, since the initial state also belongs to the fully symmetric subspace.
For this reason we have been able to simulate the dynamics up to $N=300$.

\section{Properties of the conditional states}
\label{a:conditional_states}
\subsection{Distribution of trajectory-dependent quantities}
\label{a:distributions}

\begin{figure}[h]
    \centering
    \includegraphics{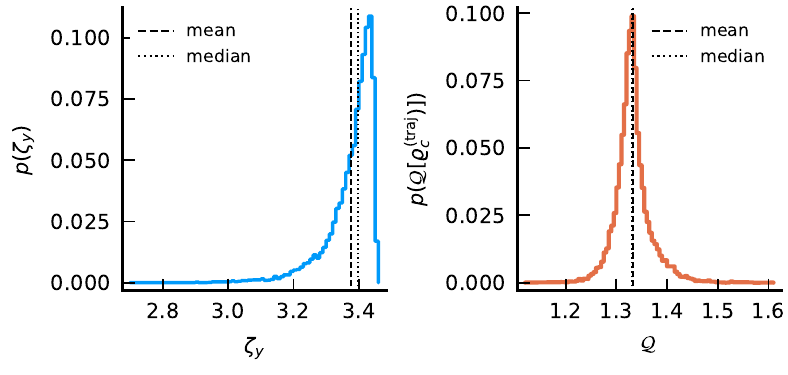}
    \caption{
    (Left) histogram of the probability density $p(\zeta_y)$ of the spin squeezing at the time  where its average $\bar\zeta_y$ is maximum ($\Gamma t\approx 0.12$).
    (Right) the distribution of $\mathcal{Q}[\varrho_c^{(\mathsf{traj})}]$ at the time where the effective QFI is maximum ($\Gamma t\approx 0.26$).
    In both plots, the black dashed (dotted) lines show the mean (median) of the distribution.
    These plots are obtained with the same parameters of Fig. 2 in the main text, for $N = 50$; in particular the histograms are obtained from $15000$ trajectories.
    }
    \label{fig:distributions}
\end{figure}

Figure 2 in the main text shows how the averages of the QFI and the spin squeezing evolve in time.
Here we look in more detail at the distribution of such quantities, considering histograms on a large number of trajectories and using the same parameters as in Fig.~2.
The left panel of Fig.~\ref{fig:distributions} shows how the spin squeezing $\zeta_y$ is distributed around its average value in the point where it is maximum for $N = 50$, while the right panel shows the distribution of the QFI.

We can notice that the spin squeezing distribution is highly asymmetric, most of the trajectories having higher squeezing than the average. The distribution of $\mathcal{Q}[\varrho_c^{(\mathsf{traj})}]$ is instead quite symmetric and peaked around the mean value. The qualitative features of these distributions do not depend on $N$.

%%%
\subsection{Visualization of the conditional state}
\label{a:visualization_conditional_state}

\begin{figure*}
    \centering
    \includegraphics[width=.9\textwidth]{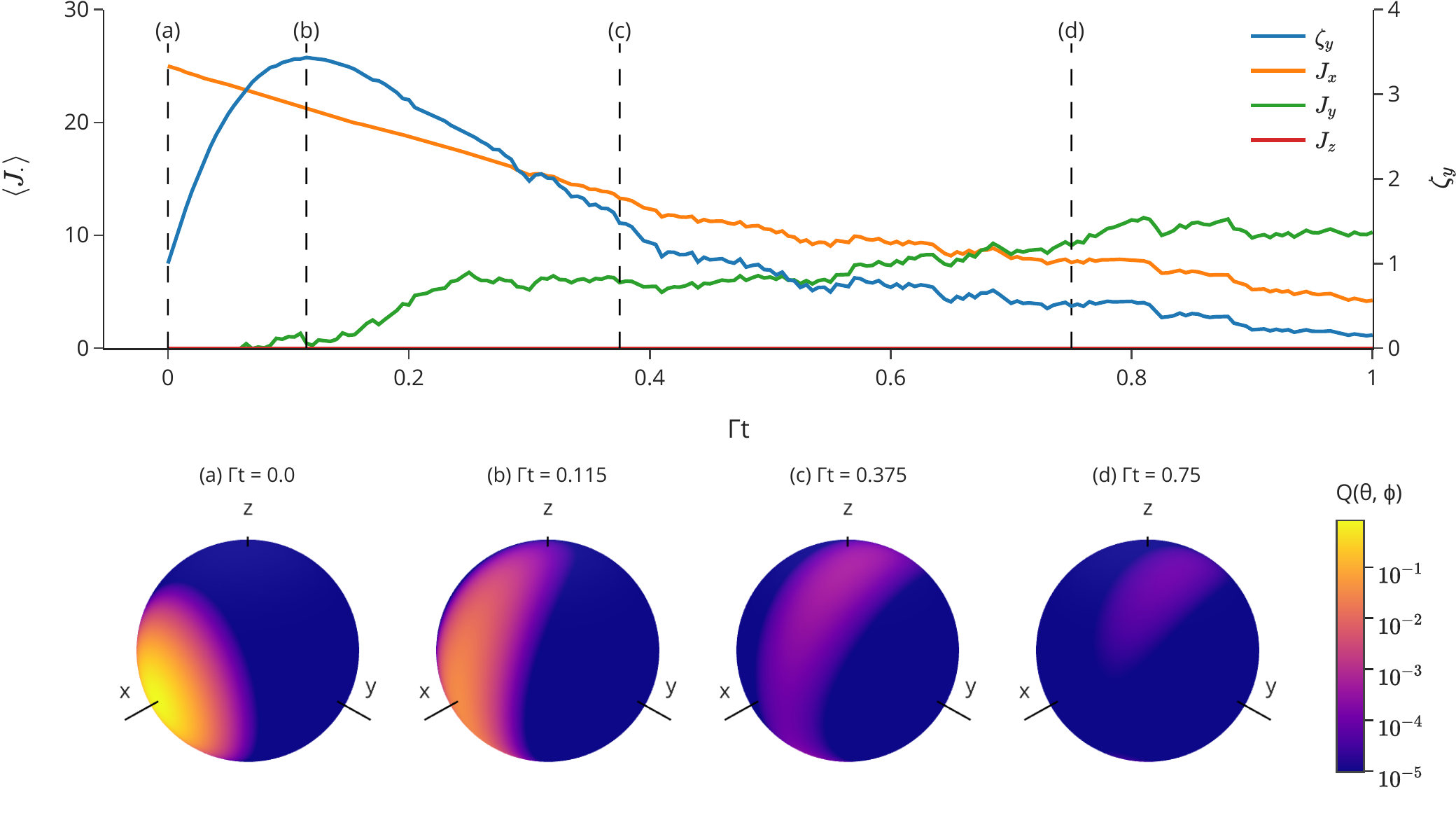}
    \caption{A single trajectory for the conditional state of a 50-qubit system, with $\kappa=\Gamma=1$, $\omega = 0.01$, $\eta=1$. The upper panel shows, on the left axis, the expectation values of $J_x$, $J_y$, and $J_z$, and on the right axis, the spin-squeezing parameter $\zeta_y$. In the lower panel, the Husimi spin Q function evaluated via Eq. \eqref{eq:spin_q_function} is plotted on the Bloch sphere at four time instants, corresponding to the vertical dashed lines in the upper panel. The initial CSS along $x$ (a) gets squeezed along the $y$ direction. The local dephasing populates the sectors with smaller total angular momentum number and thus the average value of $Q(\theta, \phi)$ decreases.}
    \label{fig:single_trajectory}
\end{figure*}

In this section we look more closely to the dynamics of the conditional states. We can visualize the state of a $N$-spin system on a Bloch sphere by using the Husimi Q function, which is defined as \cite{Agarwal1981}
\begin{equation}\label{eq:spin_q_function}
    Q(\theta, \phi) = \braket{\theta, \phi| \rho | \theta, \phi}
\end{equation}
where, by calling the maximum total angular momentum $J=N/2$, the coherent spins states can be written as
\begin{align}
    \ket{\theta, \phi} ={}& \sum_{m=-J}^{J} \binom{2J}{m + J}^{1/2}
    \left(\sin \frac \theta 2\right)^{J+m} \left(\cos \frac \theta 2\right)^{J-m} \notag \\
            & e^{-i(J+m)\phi} \ket{J, m}, \label{eq:css}
\end{align}

Figure~\ref{fig:single_trajectory} shows the dynamics of the conditional state $\rho_c$ for a single trajectory. In the upper panel, the expectation values of the components of the total momentum are shown, together with the squeezing parameter. The lower panels show the Q function,
Eq. \eqref{eq:spin_q_function}, on the Bloch sphere for four different time instants (here, $\Gamma = 1$).
%%%
\section{Finite monitoring efficiency}
\label{a:finite_efficiency}
\begin{figure}
    \centering
    \includegraphics[width=.95\columnwidth]{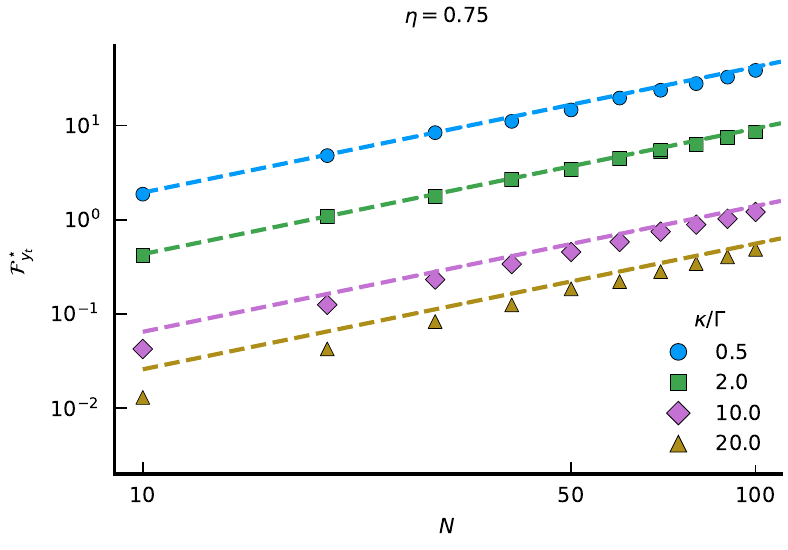} \\
    \includegraphics[width=.95\columnwidth]{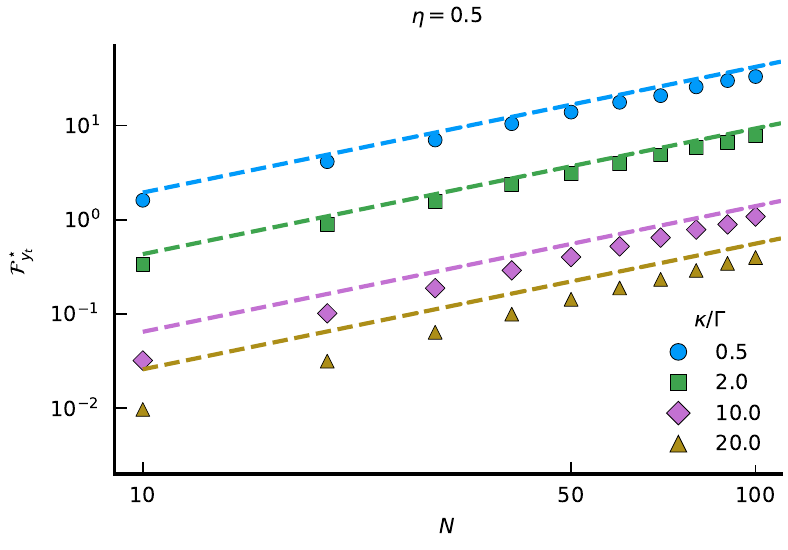}
    \caption{Optimized Fisher information for the monitoring $\monFI$ as a function of $N$ for various values of $\kappa$ for finite efficiency $\eta = 0.75$ (top panel) and $\eta = 0.5$ (bottom panel). The remaining parameters are the same as Fig.~5 of the main text.}
    \label{fig:inefficient_measurement}
\end{figure}

As discussed in the main text, a reduced measurement efficiency $\eta < 1$ for the continuous monitoring has the same qualitative effect as considering a larger dephasing. Interestingly, the scaling $N^{4/3}$ is preserved for $\monFI$, although it is achieved for larger $N$, and with a reduced proportionality constant. Figure~\ref{fig:inefficient_measurement} shows $\monFI$ as a function of $N$ for various dephasing rates for $\eta = 0.75$ (left panel) and $\eta = 0.5$ (right panel).
%
%%%
\section{Collective dephasing noise}
\label{a:collective_dephasing}

\begin{figure}
    \centering
    \includegraphics{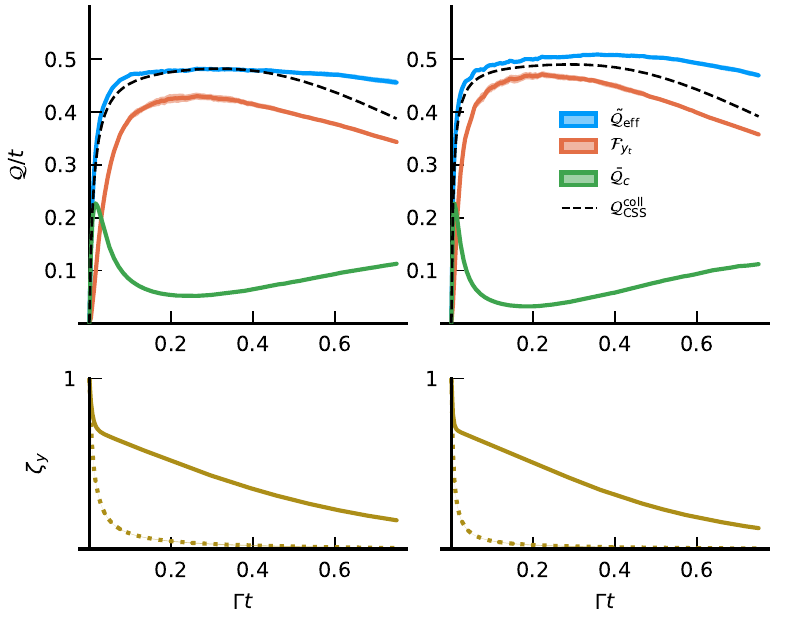}
    \caption{Top: Information rate $\mathcal{Q}/t$ for noisy frequency estimation with collective dephasing as a function of time in terms of different figures of merit, for $N = 50$ (left panels) and $N = 100$ (right panels).
Blue line: effective QFI $\effQFI/t$; orange line: continuous monitoring classical FI $\monFI/t$; green line: conditional states average QFI $\condQFI$; dashed black line: QFI $\mathcal{Q}_{\sf{CSS}}^{\sf{coll}}/t$ for a CSS affected by collective dephasing in absence of monitoring.
Bottom: average spin squeezing $\bar{\zeta}_y$ as a function of time $\Gamma t$ (dashed line is for the non-monitored dynamics).
Parameters: $\kappa_\mathsf{coll}/\Gamma=1$, $\omega/\Gamma=10^{-2}$, $\eta=1$. The shaded areas represent the $95\%$ confidence interval. The data is obtained from $20\,000$ trajectories.}
    \label{fig:QFIvsTime_coll}
\end{figure}

\begin{figure}
    \centering
    \includegraphics[width=.95\columnwidth]{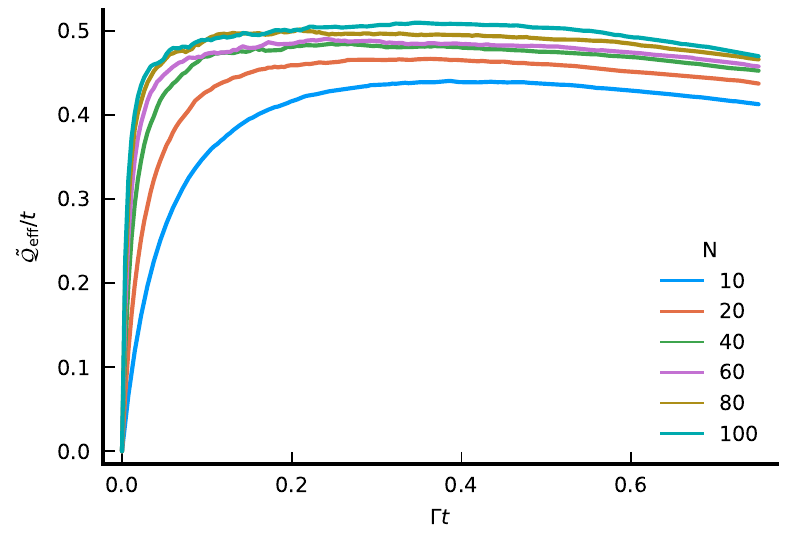}
    \caption{Effective QFI $\effQFI/t$ as a function of time $\Gamma t$ for $N$ atoms affected by collective dephasing with $\kappa_\mathsf{coll} / \Gamma = 1$. The data is obtained from $20\,000$ trajectories.}
    \label{fig:scalingQFI_coll}
\end{figure}

The discussion in the main text focuses on the effects of our QND measurement scheme in the case of local dephasing.
Here, we consider the case of collective dephasing noise, i.e. when the Lindbladian in Eq. (1) is replaced by
\begin{equation}
    \mathcal{L}\rho = -i \omega [J_z, \rho] + \kappa_{\mathsf{coll}} \mathcal{D}[J_z]\rho.
\end{equation}
If $\kappa_{\mathsf{coll}} = 2 \kappa$, in the absence of monitoring ($\Gamma=0$), the exponential decay of $\braket{J_x}$ is identical to the case of local dephasing.

Fig.~\ref{fig:QFIvsTime_coll} shows the time dependence of the metrological figures of merit already studied in Fig.~2 (main text) for independent dephasing with similar parameters.
The values of $\effQFI$ and $\monFI$ are much smaller than in the case of local dephasing and furthermore they are not increasing functions of $N$ but saturate to a constant, as shown clearly in Fig.~\ref{fig:scalingQFI_coll}.
The lower panel of Fig.~\ref{fig:QFIvsTime_coll} also shows that no spin squeezing is created during the dynamics, since it would be observed for $\xi_y > 1$.

The extremely detrimental effect of collective dephasing on standard metrological strategies is well-known~\cite{de_Falco_2013,Dorner2012a} and  we essentially show its effect also in the presence of continuous QND monitoring.
However, a fine comparison of the solid blue line with the dashed black line ($\mathcal{Q}_{\sf{CSS}}^{\sf{coll}}/t$ i.e. the QFI for a CSS affected by collective dephasing in absence of monitoring) shows that our protocol gives a (small) improvement, as mentioned in the main text.
We remark once more that this kind of noise allows to exploit different schemes based on decoherence-free subspaces ~\cite{Dorner2012a} to obtain Heisenberg scaling.
Hybrid strategies exploiting both continuous QND measurements and decoherence-free subspaces can thus be envisioned, but we do not pursue this goal in this work.

\end{document}